\documentclass{webofc}

\usepackage{mathrsfs, amsmath, amssymb, amsthm, amsxtra, bm}
\usepackage[varg]{txfonts}   
\usepackage{color}
\def\mnras{Mon. Notices Royal Astron. Soc.}             
\newcommand{\thetas}{$\theta_{\rm s}$}
\newcommand{\Ytot}{$Y_{\rm tot}$}

\newcommand{\Tsz}{$T_{\rm SZ}$}

\begin{document}
\title{Temperature measurements with the relativistic Sunyaev-Zel'dovich effect}
%
%

\author{\lastname{Y. Perrott}\inst{1}\fnsep\thanks{\email{yvette.perrott@vuw.ac.nz}} 
}

\institute{Victoria University of Wellington, New Zealand
}

\abstract{%
  At temperatures above $\sim$5\,keV, the non-relativistic approximation used to derive the classical thermal Sunyaev-Zel'dovich effect spectrum begins to fail. When relativistic effects are included, the spectrum becomes temperature-dependent. This leads to both a problem and an opportunity: a problem, because when the temperature dependence is not accounted for the Compton-$y$ estimate is biased; and an opportunity, because it represents a new way to measure the temperature of the intracluster medium independently of X-ray observations. This work presents current results from investigating the impact of relativistic effects on \emph{Planck} cluster observations, and projections for future measurements of cluster temperatures using the Atacama Large Aperture Sub-millimetre Telescope.
}
\maketitle
\section{Introduction}
\label{intro}
\begin{figure}[h]
    \centering
    \includegraphics[width=0.7\linewidth]{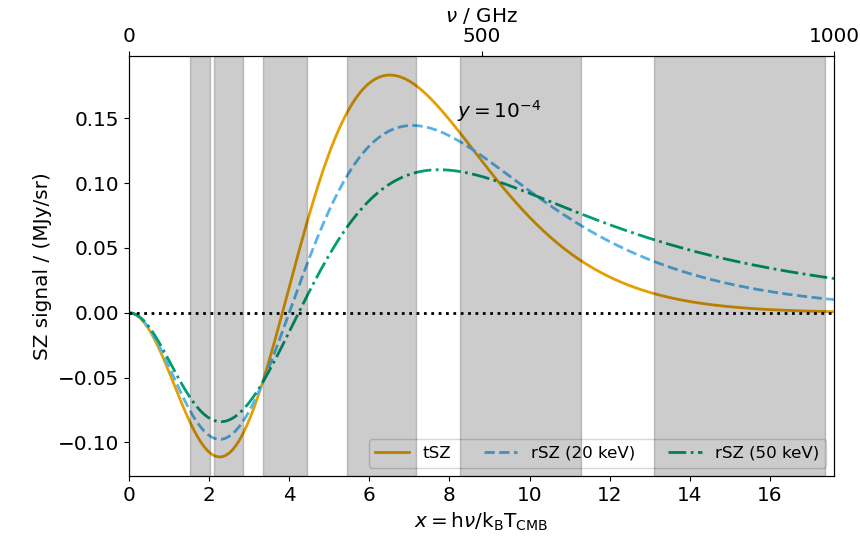}
    \caption{Change in the rSZ spectrum with increasing temperature, compared to the non-relativistic approximation (tSZ).  Grey bands show \emph{Planck} frequency bands.}
    \label{Fi:rSZ_spectrum}
\end{figure}

The standard thermal SZ spectrum is a non-relativistic limit, however it has been shown that at temperatures $\gtrapprox\,5$\,keV (typical for massive clusters), electrons start to move at non-negligible fractions of the speed of light and therefore higher-order, temperature-dependent corrections become necessary \cite{1998ApJ...499....1C, 1998ApJ...502....7I, 1998ApJ...508....1S}.  This is known as the relativistic SZ (rSZ) effect.  The rSZ frequency spectrum is temperature-dependent (see figure~\ref{Fi:rSZ_spectrum}): at frequencies less than $\approx$\,500\,GHz, the main change is a decrease in (absolute) amplitude of the signal with temperature, meaning that without strong constraints from higher frequency bands, there is a degeneracy between temperature and signal strength.  When the non-relativistic approximation is assumed, the effect is an underestimate of the overall cluster signal, which we refer to as the rSZ bias.  \cite{2018MNRAS.476.3360E} detected the relativistic correction to the SZ spectrum in a stacked sample of \emph{Planck} clusters at $\approx$\,$2\sigma$ level and predicted an rSZ bias of up to 14\% in the integrated Compton-$y$ parameter for the most massive clusters; \cite{2019MNRAS.483.3459R} considered the effect of the rSZ spectrum on the power spectrum of the Compton-$y$ parameter and concluded it could shift the constraint on $\sigma_8$ by $\approx$\,$1\sigma$, partially alleviating the tension with $\sigma_8$ measurements from the \emph{Planck} primary CMB anisotropy data.

It is well-known that X-ray temperatures can be biased due to the presence of dense, cold clumps due to the density-squared weighting of the signal; the rSZ effect, in contrast, is pressure-weighted and may be less biased (eg \cite{2008MNRAS.386.2110K}).  rSZ temperature therefore offers an independent way to measure the temperature of the ICM, and perhaps to improve our understanding of the ICM astrophysics.

Computation of the rSZ effect is now tractable with software packages such as \textsc{szpack} \cite{2012MNRAS.426..510C, 2013MNRAS.430.3054C} so it has become relatively straightforward to incorporate in SZ data analysis.  Some attempts have been made to try to constrain rSZ temperatures on individual clusters (eg \cite{2022ApJ...932...55B}) although the constraints so far have been weak.

\section{rSZ with \emph{Planck}}
\label{Planck}

The \emph{Planck} frequency bands span an appropriate frequency range to distinguish between the rSZ effect and a decrease in Compton-$y$.  We investigated whether it was possible to constrain rSZ temperatures with \emph{Planck} data by creating simulations using a realistic physical model for the cluster thermodynamic profiles \cite{2012MNRAS.423.1534O, 2013MNRAS.430.1344O}, then generating rSZ signal images using \textsc{szpack}.  The simulated clusters were injected into real \emph{Planck} data in areas away from real clusters, and the resulting images analyzed using a version of the \textsc{powellsnakes} software \cite{2009MNRAS.393..681C, 2012MNRAS.427.1384C} adapted to also fit for a single average temperature.  We used a simple cluster parameterisation defined by the characteristic angular scale of the cluster (\thetas) and its total integrated Compton-$y$ parameter (\Ytot) to fit the simulated data.

\begin{figure}[h]
    \centering
    \includegraphics[width=0.8\linewidth]{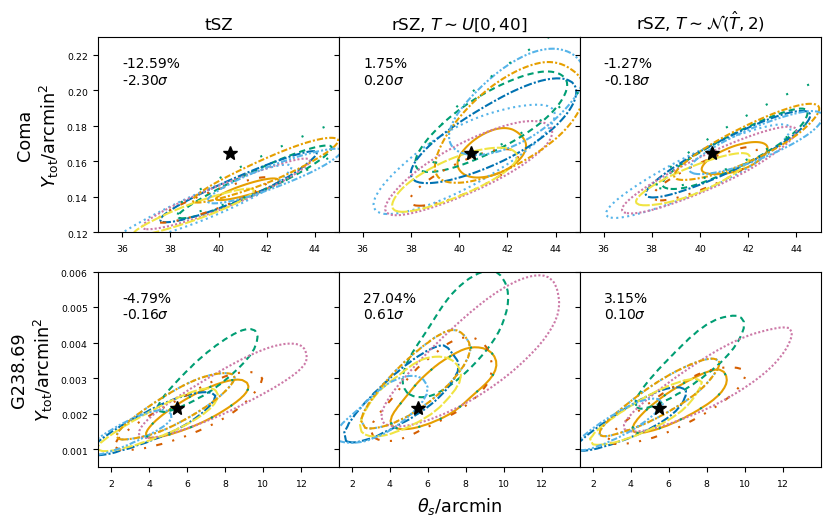}
    \caption{Posterior constraints on the cluster \thetas\ and \Ytot\ model parameters derived from analysing relativistic simulations with the non-relativistic approximation to the SZ signal (left column) and isothermal relativistic signal (other columns; prior on \Tsz\ given in title).  For a given cluster, all plots have the same axis ranges and the true value is marked with a black star.  The 68\% probability contour is shown with different colours/line styles for a random selection of ten realizations of the simulation.  Average 1D \Ytot\ bias values for the whole simulation set are given on each figure, both as a percentage and as an average significance level.}
    \label{Fi:sim_posteriors}
\end{figure}

In figure~\ref{Fi:sim_posteriors} we show example results from some realizations of two of the simulated clusters.  The first is a high-SNR cluster with parameters chosen to be similar to Coma, while the second is more representative of the median of the \emph{Planck} population.  When assuming the tSZ spectrum (left-hand column), the recovered \Ytot\ values are significantly offset from the true value for the Coma-like cluster, by an average of $\approx$\,13\%.  For the lower-SNR cluster, the bias is not significant on the level of an individual realization, but over the whole simulation population \Ytot\ is biased down by $\approx$\,5\%.

We attempted to constrain the rSZ temperature by putting a uniform prior on temperature in the analysis (middle column of figure~\ref{Fi:sim_posteriors}).  In the case of Coma, this does successfully produce a (weak) constraint on temperature and an unbiased \Ytot\ value.  However, in the case of the low-SNR cluster, the temperature is not constrained and the degeneracy between \Ytot\ and temperature leads to a large \textit{upward} bias in \Ytot.  We therefore conclude that at the sensitivity level of \emph{Planck}, for most clusters it will not be possible to constrain temperature and attempting to do so will lead to \Ytot\ constraints that are biased up rather than down.  We can, however, avoid the downward rSZ bias by putting an informative prior on temperature (right-hand column of figure~\ref{Fi:sim_posteriors}) such as a Gaussian centred at the true value.  In the case of the simulations, the `true' (pressure-weighted average) value is known; in the case of real data an external estimate of the temperature from X-ray could be used, although care should be taken in considering the different weightings involved in averaging over the cluster volumne.  Alternatively, a scaling relation between, eg, \Ytot\ and temperature such as those calibrated from numerical simulations by \cite{2020MNRAS.493.3274L, 2022MNRAS.517.5303L} can also be used.

Although these biases are small and not necessarily significant for individual clusters, when considering the population as a whole (i.e. for constructing scaling relations), the rSZ bias produces a mass-dependent decrease in \Ytot\ which skews the scaling relation.  When calibrating a scaling relation \emph{on} \emph{Planck} data for applying \emph{to} \emph{Planck} data the bias cancels out; however when comparing scaling relations between SZ measurements made by different instruments and/or simulations, this can have a significant effect.  A corrected \emph{Planck} scaling relation will be presented in a forthcoming publication (Perrott et al, in prep).

\section{rSZ with AtLAST}
\label{AtLAST}

The Atacama Large Aperture Sub-millimetre Telescope (AtLAST; \cite{2022SPIE12190E..07R}) is a proposed 50m single dish telescope to be situated in the Atacama desert, allowing the high-frequency observations necessary for constraining rSZ temperature.  Its high angular resolution will allow both spatial separation of confusing foreground/background emission, and the possibility to resolve SZ effect signals.

We assume the bands in table~\ref{T:AtLAST_bands} to make the following predictions, and the current sensitivity calculator estimates\footnote{\url{https://github.com/lucadimascolo/AtLAST_sensitivity_calculator.git}}.  The band structure and sensitivity calculator are under development so these estimates should be taken as preliminary only.

\begin{table}
\centering
\caption{Bands assumed to make rSZ predictions for AtLAST.}
\label{T:AtLAST_bands}       
\begin{tabular}{cc}
\hline
Lower limit / GHz & Upper limit / GHz  \\\hline
30 & 54 \\
66 & 117 \\
120 & 182 \\
183 & 252 \\
252 & 325 \\
325 & 375 \\
384 & 422 \\
595 & 713 \\
786 & 905 \\
\hline
\end{tabular}
\end{table}

\begin{figure}[h]
  \centering
    \includegraphics[width=0.49\linewidth]{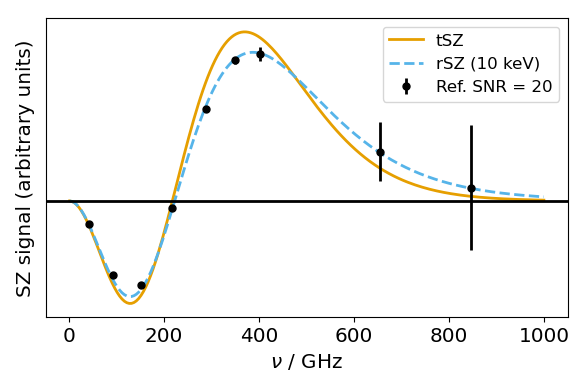}
    \includegraphics[width=0.49\linewidth]{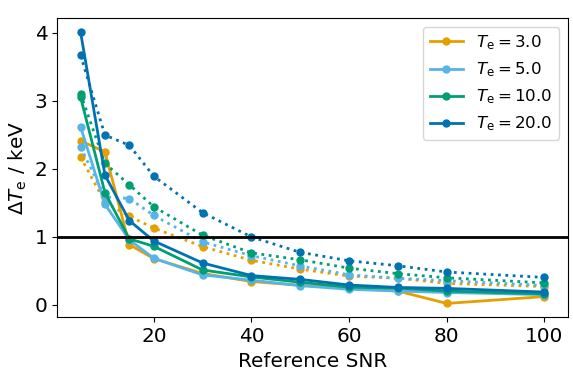}
    \caption{Left: simulated AtLAST observation of a $T_\mathrm{e}=10$\,keV cluster, with SNR=20 in the reference band.  Right: rSZ temperature 68\% confidence limits as a function of reference band SNR for different input temperatures.  Dotted (solid) lines show the case that there is (not) a correlated dust component.  The horizontal black line shows a desirable constraint level of $1$\,keV, which would be reached by fitting to the spectrum on the left.}
    \label{Fi:AtLAST_SNR}
\end{figure}

We first ask the question, what sensitivity is required to constrain rSZ temperature?  For a given temperature, we generate an rSZ spectrum as in Fig.~\ref{Fi:rSZ_spectrum} and average within the bandpasses shown in Table~\ref{T:AtLAST_bands} to simulate an AtLAST observation.  We assume the same observing time for all bands to generate error estimates in each band, taking into account the variation both in point source sensitivity and in beam sizes amongst the frequency bands, i.e.

\begin{equation}
\frac{\Delta I(n)}{\Delta I(\mathrm{ref})} = \frac{\sigma(n)}{\sigma(\mathrm{ref})} \times \frac{\Omega(\mathrm{ref})}{\Omega(n)},
\end{equation}
where $I$ is the SZ signal in MJy\,sr$^{-1}$, $n$ denotes the band index, and $\sigma(n)$ and $\Omega(n)$ are the point source flux density sensitivity and beam size in a given band $n$, respectively.  An example simulated spectrum is shown in Fig.~\ref{Fi:AtLAST_SNR} (left).

We then fit the simulated flux density measurements for $y$ and $T_\mathrm{e}$ to test the error on the recovered $T_\mathrm{e}$ estimate.  By doing this for a range of temperatures and observing times, we can test how well temperature is constrained as a function of SNR in a reference band, chosen to be the $\approx$\,400\,GHz band.  The results are shown in figure~\ref{Fi:AtLAST_SNR}; we conclude a reference SNR of 40 is sufficient for a $<1$\,keV temperature constraint in all cases, and a reference SNR of 20 is sufficient for a $<1$\,keV temperature constraint if there is no correlated dust component, or $<2$\,keV temperature constraint if there is a spatially correlated dust component which must be separated spectrally, as detected by, eg \cite{2018MNRAS.476.3360E}.

\begin{figure}[h]
    \centering
    \includegraphics[width=0.49\linewidth]{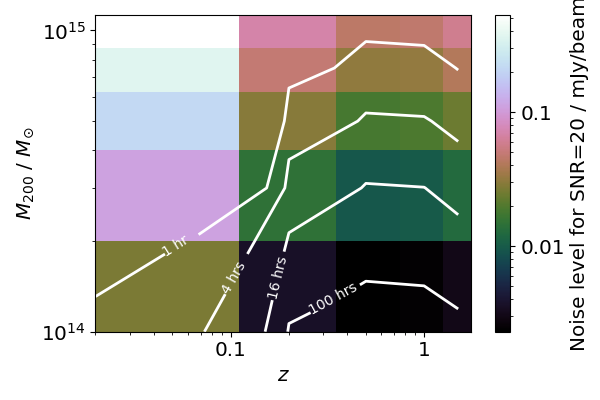}
    \includegraphics[width=0.49\linewidth]{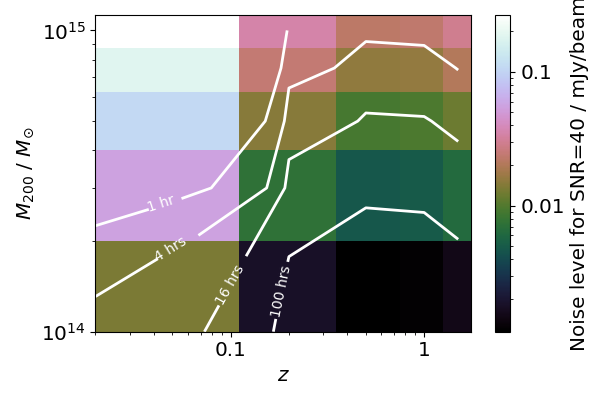}
    \caption{Noise levels, and corresponding predicted observing times, required to get to a reference band SNR = 20 (left) and 40 (right) with AtLAST.}
    \label{Fi:AtLAST_observing_times}
\end{figure}

Given these results, we next predict what observing time will be needed to obtain the required SNR for a global temperature estimate, using a simple physically motivated model to create signal maps as in Section~\ref{Planck}.  Based on the signal maps, we estimate the cylindrically-integrated signal within $r_{200}$ in the reference band and use the sensitivity calculator to estimate the observing time required to reach the required SNR of 20 or 40 for a given cluster mass and redshift.  These results are shown in figure~\ref{Fi:AtLAST_observing_times}.  We see that with modest amounts of observing time, rSZ constraints should be possible for all low-redshift ($z<0.1$) clusters, and clusters with $M_{200}\gtrapprox 3\times 10^{14} M_\odot$ across the whole redshift range.

\begin{figure}[h]
    \centering
    \includegraphics[width=0.49\linewidth]{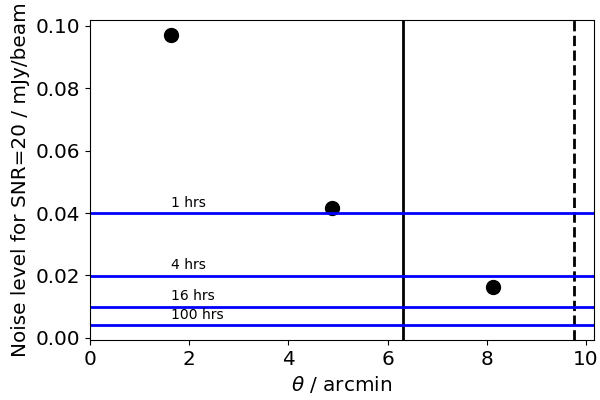}
    \includegraphics[width=0.49\linewidth]{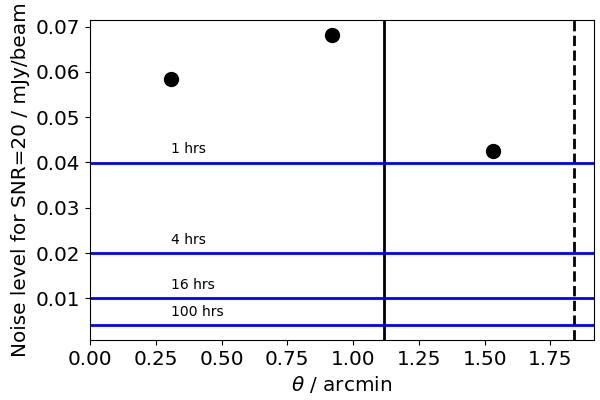}
    \caption{Noise levels required to get to a reference band SNR = 20 in angular bins for a cluster of $M_{200}=10^{15} M_\odot$ at $z=0.2$ (left) and $M_{200}=5\times 10^{14} M_\odot$ at $z=1.5$ (right). Predicted observing times are shown with horizontal blue lines, and $\theta_{500}$ and $\theta_{200}$ with vertical black solid and dashed lines respectively.}
    \label{Fi:AtLAST_resolved}
\end{figure}

Given AtLAST's angular resolution, we would also like to test whether resolved rSZ temperature profiles will be possible.  To do this, we divide each cluster simulation into three annular bins inside $\theta_{200}$ and check the observing time required to reach an SNR of 20 in the reference band, with respect to the signal in each annulus.  Two example results are shown in figure~\ref{Fi:AtLAST_resolved}.  For a high mass cluster of $M_{200}=10^{15} M_\odot$ at $z=0.2$, resolved temperatures are constrained within $\theta_{500}$ after 4 hours observing time.  For a lower-mass cluster of $M_{200}=5\times 10^{14} M_\odot$ at $z=1.5$, resolved temperatures are constrained within $\theta_{200}$ after 1 hour observing time.  The improvement at high redshift (also seen in figure~\ref{Fi:AtLAST_observing_times}) is likely due to the angular diameter distance turnover.

\section{Conclusion}

With the precision and sensitivity of current and forthcoming instruments, the non-relativistic SZ spectrum is no longer an adequate approximation.  With current instruments such as \emph{Planck}, this mainly presents a problem in that Compton-$y$ estimates are biased when it is neglected, but sensitivity limits (\emph{Planck}) and/or lack of high-frequency channels (ACT, SPT) make it difficult to constrain.  If external temperature estimates are used, for example from X-ray measurements or scaling relations calibrated using numerical simulations, the bias can be accounted for and removed.  This is particularly necessary when comparing SZ scaling relations between different instruments and/or with numerical simulations.

Forthcoming instruments with higher sensitivity and the required high-frequency capability (AtLAST, FYST) will be able to do better.  Simulations suggest that AtLAST will be able to measure global and resolved rSZ temperature estimates in reasonable amounts of observing time.  For these instruments, the rSZ effect becomes an opportunity since it offers an alternative way to measure temperature that simulations suggest should be less biased than X-ray temperature measurements.  This offers an exciting new avenue for probing cluster astrophysics in the future.

%
%

\end{document}